\documentclass{ws-mpla}

\begin{document}

\markboth{Ke-Sheng Sun,etc.}
{LEPTON FLAVOR VIOLATION IN INVERSE SEESAW MODEL}

\catchline{}{}{}{}{}

\title{LEPTON FLAVOR VIOLATION IN INVERSE SEESAW MODEL}

\author{KE-SHENG SUN    $^{a,b,\ast}$ ,
        TAI-FU FENG     $^{b}$,
        GUO-HUI LUO     $^{a}$,
        XIU-YI YANG     $^{c,d}$,\\
        JIAN-BIN CHEN   $^{e}$}

\address{$^{a}$   Department of Physics, Dalian University of Technology, Dalian 116024, China\\
         $^{b}$   Department of Physics, Hebei University, Baoding 071002,China\\
         $^{c}$   School of Physics, Shandong University, Jinan 250100,China\\
         $^{d}$   School of Science, University of Science and Technology Liaoning, Anshan 114051,China\\
         $^{e}$   College of Physics and Optoelectronics, Taiyuan University of Technology, Taiyuan 030024, China\\
         $^{\ast}   $    sunkesheng@126.com}

\maketitle

\pub{Received (Day Month Year)}{Revised (Day Month Year)}

\begin{abstract}
We analyze the lepton flavor violation processes $\mu-e$ conversion, $l_i\rightarrow l_j\gamma$ and $l_i\rightarrow 3l_j$ in framework of the Standard Model (SM) extended with inverse seesaw mechanism, as a function of $\tilde{\eta} = 1 -|Det(\tilde {U}_{PMNS})|$ that parameterizes the departure from unitary of the light neutrino mixing sub-matrix $\tilde {U}_{PMNS}$. In a wide range of $\tilde{\eta}$, the predictions on the $\mu-e$ conversion rates and the branching ratio of $\mu\rightarrow e\gamma$ are sizeable to be compatible with the experimental upper limits or future experimental sensitivities. For large scale of $\tilde{\eta}$, the predictions on branching ratios of other lepton flavor processes can also be reach the experimental upper limits or future experimental sensitivities. The value of $\tilde{\eta}$ depends on the determinant of the Majorana mass term $M_{\mu}$. Finally, searching for lepton flavor violation processes in experiment provides us more opportunities for the searches of seesaw nature of the neutrino masses.
\end{abstract}

\section{Introduction}\label{intro}	
Neutrino oscillation experiments \cite{Fukuda1,Fukuda2,Ahmad,Eguchi} have established compelling evidence that neutrinos are massive and the neutral lepton
flavor is not conserved. In SM extended with massive neutrinos, the charged Lepton Flavor Violation (LFV) processes, arising from loop level, such as radiative two body decays $(l_i\rightarrow l_j\gamma)$ and leptonic three body decays $(l_i\rightarrow 3l_j)$, remain highly suppressed, see Table.\ref{BRCR}\cite{PDG} for current experimental upper bounds, making them difficult to observe. The limit on branching ratio of $\mu\rightarrow e\gamma $ is the most recent result given by the MEG experiments at the 90\% confidence level \cite{MEG2}. Nevertheless, various extensions of the SM, such as the seesaw model with or without GUT, supersymmetry, $Z^{\prime}$ models, etc., have predicted enhanced branching ratios of LFV processes to be accessible in current experiments. Thus, searching for LFV processes are a powerful way to prove physics beyond the SM.

\begin{table}[h]
\tbl{Current limits and future expectations for $\mu- e$ conversion, $l_i\rightarrow l_j\gamma$ and $l_i\rightarrow 3l_j$. }
{\begin{tabular}{@{}ccc@{}}
\toprule
Channel&Limit&Future\\
\colrule
$CR(\mu-e,Al)$&-&$2.4\times10^{-17}$,\cite{Abrams},$10^{-17}$,\cite{Kuno}\\
$CR(\mu-e,Ti)$&$<4.3\times10^{-12}$&$10^{-18}$,\cite{Barlow}\\
$CR(\mu-e,SiC)$&-&$10^{-14}$,\cite{DeeMe}\\
$CR(\mu-e,Au)$&$<7\times10^{-13}$&-\\
$CR(\mu-e,Pb)$&$<4.6\times10^{-11}$&-\\
$BR(\mu\rightarrow e\gamma )$&$<5.7 \times 10^{-13}$&$6\times10^{-14}$,\cite{MEG}\\
$BR(\tau\rightarrow e\gamma)$&$<3.3 \times10^{-8}$&$2.3\times10^{-9}$,\cite{SuperB}\\
$BR(\tau\rightarrow \mu\gamma)$&$<4.4\times10^{-8}$&$3\times10^{-9}$,\cite{BelleII},$1.8\times 10^{-9}$,\cite{SuperB}\\
$BR(\mu\rightarrow 3e)$&$<1.0 \times 10^{-12}$&$10^{-15}$,\cite{Yoshida},$10^{-16}$,\cite{Blondel}\\
$BR(\tau\rightarrow 3e)$&$<2.7\times10^{-8}$&$2\times10^{-10}$,\cite{SuperB}\\
$BR(\tau\rightarrow 3\mu)$&$<2.1\times10^{-8}$&$10^{-9}$,\cite{BelleII},$2\times 10^{-10}$,\cite{SuperB}\\ \botrule
\end{tabular}\label{BRCR}}
\end{table}

The seesaw mechanisms have been recognized as the most natural scenario for understanding the smallness of neutrino mass up to now. In canonical Type(I) seesaw, three right-handed neutrinos are introduced, and to achieve sub-eV range of light neutrino masses, Grand Unified (GUT) scale (i.e., $10^{16}$ GeV) of the right-handed neutrinos is required and that makes the LHC study of the new physics scale difficult.
In order to make the right-handed neutrino masses down to the TeV scale, the small neutrino masses have to be effectively suppressed via other mechanisms rather than the GUT scale, such as radiative generation, small lepton number breaking, or neutrino masses from a higher than dimension-five effective operator \cite{Bonnet}.
Another option to relate small neutrino mass to TeV scale physics is the inverse seesaw mechanism \cite{Mohapatra1,Mohapatra2}. The smallness of the light neutrino masses can be ascribed to the smallness of $M_{\mu}$, which breaks the lepton number by two unity.

The smallness of $M_{\mu}$ is a key element of the inverse seesaw models. So far, a very appealing picture is the radiative origin of the
two unity lepton number-breaking parameter as it has been proposed in Ref.\refcite{Ma}: it is induced at two-loop level, thus
explaining its smallness with respect to the electroweak scale (EW). Introducing new scalar fields, the two unity lepton
number-breaking term can also be induced at two-loop level and is naturally around the keV scale, while righthanded
neutrinos are at the TeV scale \cite{Bazzocchi}. In the supersymmetric inverse seesaw mechanism, the smallness of $M_{\mu}$ was
related to vanishing trilinear susy soft terms at the grand unified theory (GUT) scale \cite{Bazzocchi1}. In warped extra dimension, one can have the $M_{\mu}$ smallness dictated by parameters of order one that govern the location of the 5D profile of the S fields in the bulk \cite{Fong}.

The effective mass matrix for the light neutrinos is given by
\begin{eqnarray}
m_{\nu} \approx M_D^T (M^T_R)^{-1} M_{\mu} (M_R)^{-1} M_D.
\end{eqnarray}
So that scale of $M_R$ can be made small and many phenomena due to the non-unitary feature of the neutrino mixing matrix can be manifested, such as LFV, CP violation and non standard effects in neutrino propagation \cite{Forero}. Non-unitary mixing between light and heavy particles can be large and can be probed at colliders \cite{Bandyopadhyay1,Das,DeV,Bandyopadhyay2} and future neutrino factories \cite{Fernandez,Goswami,Antusch1}.

The LFV effect of the non-unitary feature of the neutrino mixing matrix have been investigated in several literatures where the inverse seesaw mechanism is accommodated in SM \cite{Forero,Ilakovac,Antusch,Alonso,Deppisch}, SM with B - L extension \cite{Abdallah} and supersymmetric models \cite{Deppisch2,Abada1,Abada2,Bhupal}. It is shown that LFV decay $\mu\rightarrow e\gamma$ can be sizeable to be observed in experiment, where the scale of $M_R$ is fixed to $10^{3}$ GeV and the scale of $M_{\mu}$ varies in the range of $[10^{-10},10^{-8}]$ GeV, and to be compatible with the experiment limit on $\mu\rightarrow e\gamma$, large value setup of $M_{\mu}$ is favored \cite{Forero}. Assuming $\Delta L = 2$ interactions are absent from
the model, i.e.,$M_{\mu}=0$, Ref.\refcite{Ilakovac} estimate the $BR(\tau\rightarrow 3e)$ or $BR(\tau\rightarrow e\mu\mu)$ can be large as $10^{-6}$ and the limits are out of date.
In the inverse seesaw model, the limits on degenerate values of $M_R$ and $M_{\mu}$ from the photonic contribution are much more stringent than from the non-photonic contribution for $\mu-e$ conversion in nucleus, and the rates arising from virtual photon exchange are generically correlated to the $\mu\rightarrow e\gamma$ decay \cite{Deppisch}. It is also shown that prediction on the branching ratio of $\mu\rightarrow e\gamma$ can be within the reach of MEG experiment in B - L extension of the SM with inverse seesaw mechanism \cite{Abdallah}. In supersymmetric inverse seesaw model, the LFV  decays can be enhanced by flavour violating slepton contributions, the non-unitary of the charged current mixing matrix or the Higgs mediated processes \cite{Deppisch2,Abada1,Abada2}. In the framework of a supersymmetric $SO(10)$ model with inverse seesaw \cite{Bhupal}, the expected branching ratio for $(l_i\rightarrow l_j\gamma)$ are several orders of magnitude below the future sensitivity in experiment with TeV scale slepton mass, and for $(l_i\rightarrow 3l_j)$ and $\mu - e$ conversion, the predictions are much smaller than what can be probed in planned experiments.

In SM, the LFV decays mainly originate from the charged current with the mixing among three lepton generations.
The fields of the flavor neutrinos in charged current weak interaction Lagrangian are combinations of three massive neutrinos:
\begin{eqnarray}
{\cal L}& = &-\frac{g_2}{\sqrt{2}}\sum_{l=e,\mu,\tau}\overline{l_{L}}(x)\gamma_\mu \nu_{lL}(x)W^\mu(x)+h.c.,
\nonumber\\
\nu_{lL}(x)& = &\sum_{i=1}^{3}\Big(U_{PMNS}\Big)_{li}\nu_{iL}(x),
\end{eqnarray}
where $g_2$ denotes the coupling constant of gauge group SU(2), $\nu_{lL}$ are fields of the flavor neutrinos, $\nu_{iL}$ are fields of massive neutrinos, and $U_{PMNS}$ corresponds to the unitary neutrino mixing matrix \cite{Pontecorvo1,Pontecorvo2,Maki}.

In this paper we have studied LFV decays $l_i\rightarrow l_j\gamma$, $l_i\rightarrow 3l_j$ and $\mu-e$ conversion as a function of non-unitary parameter $\tilde{\eta}$, which is firstly introduced in Ref.\refcite{Abada} in the SM extended with inverse seesaw mechanism. Moreover, we also investigate the the dependence of $\tilde{\eta}$ on $M_{\mu}$. From this point of view, the paper proposed is different from others. We perform a scan over non-degenerate parameters $M_R$ and $M_{\mu}$, which vary in region of $[1,10^6]$ GeV and $[10^{-11}, 10^{-3}]$ GeV, respectively, by taking account of the constraints from neutrino oscillation data and several rare decays. We have give a discussion about the parameter spaces, which is more narrow than Ref.\refcite{Deppisch}. For $CR(\mu-e, Nucleus)$, both photonic and non-photonic contribution are considered in this paper.

The paper is organized as follows. In Section.\ref{sec2}, we review the inverse seesaw mechanism and give the expressions for the unitary violating parameter $\tilde{\eta}$. The numerical results and discussions are presented in Section.\ref{sec3}. The conclusion is drawn in Section.\ref{sec4}.

\section{Inverse seesaw model}
\label{sec2}
The inverse seesaw mechanism can be accommodated in SM by adding two kind of singlet fermions, $N^i_R$ and $S^i_R$, and one gauge singlet scalar $\Phi$ to the SM field content, where $N^i_R$ ($i$ = 1,2,3) stand for the usual right-handed neutrinos, $S^i_R$ ($i$ = 1,2,3) stand for the additional gauge singlet neutrinos, and these two kind fermions share opposite lepton number (-1 and 1, respectively).
The relevant gauge invariant Lagrangian for neutrino masses is given by \cite{Mohapatra1,Mohapatra2,Bazzocchi,Chan}:
\begin{eqnarray}
\mathcal {L}= \overline{N^c_R}Y_{\nu} \widetilde{H}l_L+ \overline{N^c_R}Y^{\prime}_{\nu}S_R \Phi+\frac{1}{2}\overline{S^c_R}M_{\mu}S_R+h.c.,
\label{LISS}
\end{eqnarray}
where $l_L$ stands for the $SU(2)_L$ lepton doublet, $\widetilde{H}\equiv i \sigma_2 H^{\ast}$ stands for the Higgs doublets, $Y_{\nu}$ and $Y^{\prime}_{\nu}$ are the $3\times 3$ Yukawa coupling matrices, and $M_{\mu}$ is a symmetric Majorana mass matrix.
In this mechanism, it introduces an extra $U(1)$ gauge symmetry into the electroweak model, under which the right-handed neutrino must be a non-singlet. After spontaneous gauge symmetry breaking, the extra $U(1)$ gauge group breaks into $U(1)_Y$, the weak hypercharge of the standard
model. The invariant Lagrangian in Eq.(\ref{LISS}) would be:
\begin{eqnarray}
\mathcal {L}= \overline{\nu_L}M_D N^c_R + \overline{N^c_R}M_R S_R +\frac{1}{2}\overline{S^c_R}M_{\mu}S_R+h.c.,
\label{LISSSGSB}
\end{eqnarray}
where $M_D = Y_{\nu}\langle H\rangle = \frac{\upsilon}{\sqrt{2}}Y_{\nu}$ and $M_R = Y^{\prime}_{\nu}\langle \Phi\rangle$ are $3\times3$ mass matrices, with $\upsilon$ the vacuum expectation value of the SM Higgs boson. It shows that the right-handed neutrino mass term $M_R$ conserves lepton number and the Majorana mass term $M_{\mu}$ violates the lepton number by two units.

The neutrino mass matrix in the flavor basis defined by $(\nu_L, N^c_R, S^c_R)$ is given by
\begin{eqnarray}
\mathcal{M}=
\left(\begin{array}{ccc}
0&M^T_D&0\\
M_D&0&M_R\\
0&M^T_R&M_{\mu}
\end{array}\right),
\label{Mass}
\end{eqnarray}
where $\mathcal{M}$ is a $9\times 9$ matrix. The mass scales of $M_D$,$M_R$ and $M_{\mu}$ in Eq.(\ref{Mass}) may naturally have a hierarchy $M_R\gg M_D\gg M_{\mu}$ \cite{Hooft}. In reality, a small non-vanishing $M_{\mu}$ can be viewed as a
slight breaking of a global $U(1)$ symmetry.
The $9\times 9$ neutrino mass matrix $\mathcal{M}$ can be diagonalized by the unitary mixing matrix $U$:
\begin{eqnarray}
\hat{\mathcal{M}}= U^{T}\mathcal{M}\hspace{1mm}U,
\label{U}
\end{eqnarray}
and it yields nine mass eigenstates $N_i$.
The light neutrino flavour states $\nu_{lL}$ could be given in terms of the mass eigenstates via the unitary
matrix $U$ as
\begin{eqnarray}
\nu_{lL} = \sum^9_{i = 1} (U)_{l i}N_{i}.
\end{eqnarray}
It is obvious that the mixing matrix would be simply the rectangular matrix formed by the first three rows of $U$ in Eq.(\ref{U}) and the matrix $\tilde {U}_{PMNS}$ describing the mixing between the charged leptons and light neutrinos in inverse seesaw mechanism could be written by:
\begin{eqnarray}
\tilde {U}_{PMNS}
&=&\left(\begin{array}{ccc}
U_{11}&U_{12}&U_{13}\\
U_{21}&U_{22}&U_{23}\\
U_{31}&U_{32}&U_{33}
\end{array}\right).
\label{TPMNS}
\end{eqnarray}
In inverse seesaw mechanism, $U$ in Eq.(\ref{U}) is unitary. However, $\tilde {U}_{PMNS}$ is not unitary. To parametrize this departure from unitary, we could define $\tilde{\eta}$ as in Ref.\refcite{Abada} by:
\begin{eqnarray}
\tilde{\eta} = 1 -|Det(\tilde {U}_{PMNS})|.
\end{eqnarray}
It has been shown in Ref.\refcite{Abada} that large value of $\tilde{\eta}$ is responsible for the lepton flavour universality violation in $K^+$ and $\pi^+$ leptonic decays in SM extended with inverse seesaw mechanism.

The diagonalization of $\mathcal {M}$ leads to an effective mass matrix for the light neutrinos in the leading order approximation \cite{Gonzalez},
\begin{eqnarray}
m_{\nu} = M_D^T (M^T_R)^{-1} M_{\mu} (M_R)^{-1} M_D,
\label{Mv}
\end{eqnarray}
which indicates that the light neutrino masses vanish in the limit $M_{\mu} \rightarrow 0$ and lepton number conservation is restored.
The effective mass matrix $m_{\nu}$ is diagonalized by the physical neutrino mixing matrix $U_{PMNS}$,
\begin{eqnarray}
\hat m_{\nu}= U^{T}_{PMNS} m_{\nu} U_{PMNS},
\end{eqnarray}
and, in the standard parametrization \cite{PDG}, $U_{PMNS}$ is given by
\begin{eqnarray}
U_{PMNS}
&=&\left(\begin{array}{ccc}
c_{1}c_{3}&c_{3}s_{1}&s_{3}e^{-i\delta}\\
-c_{1}s_{3}s_{2}e^{i\delta}-c_{2}s_{1}&c_{1}c_{2}-s_{1}s_{2}s_{3}e^{i\delta}&c_{3}s_{2}\\
s_{1}s_{2}-c_{1}s_{3}c_{2}e^{i\delta}&c_{1}s_{2}-s_{1}c_{2}s_{1}e^{i\delta}&c_{3}c_{2}
\end{array}\right)\nonumber\\
&&\times   diag\Big(e^{i\Phi_{1}/2},1,e^{i\Phi_{2}/2}\Big),
\label{MNS}
\end{eqnarray}
where $s(c)_{1}$ = $\sin(\cos)\theta_{12}$, $s(c)_{2}$ = $\sin(\cos)\theta_{23}$, $s(c)_{3}$ = $\sin(\cos)\theta_{13}$, and the experimental limits on the mixing angles are given in Table.\ref{Ndata}.
The phase $\delta$ is the Dirac CP phase, and $\Phi_{i}$ are the Majorana phases. The remaining six heavy states have masses approximately given by $M_{\nu}\simeq M_R$.

Without loss of generality, we work in a basis where $M_R$ is assumed as diagonal matrix. Using a modified Casas-Ibarra parametrisation \cite{Casas}, which is automatically reproducing the light neutrino data, $Y_{\nu}$ can be written by
\begin{eqnarray}
Y_{\nu} = \frac{\sqrt{2}}{\upsilon}V^{\dagger}\sqrt{\hat M}R\sqrt{\hat m_{\nu}} U^{\dagger}_{PMNS},
\label{Yv}
\end{eqnarray}
with $\upsilon$ the vacuum expectation value of the SM Higgs boson. $\hat M$ is the relevant diagonal matrix of $M=M_{R}M^{-1}_{\mu} M_{R}^{T}$, which is diagonalized by matrix $V$:
\begin{eqnarray}
\hat M= V M V^{T}=V M_{R}M^{-1}_{\mu} M_{R}^{T} V^{T},
\end{eqnarray}
and $R$ is a $3\times 3$ complex orthogonal matrix, parametrized by three complex angles $\alpha_{1}$, $\alpha_{2}$, $\alpha_{3}$:
\begin{eqnarray}
R
&=&\left(\begin{array}{ccc}
c_{2} c_{3} &-c_{1} s_{3}-s_{1} s_{2} c_{3}  &s_{1} s_{3}-c_{1} s_{2} c_{3}\\
c_{2} s_{3} &c_{1} c_{3}-s_{1} s_{2} s_{3}  &-s_{1} c_{3}-c_{1} s_{2} s_{3} \\
 s_{2} &s_{1} c_{2}  &c_{1} c_{2}
\end{array}\right),
\label{R}
\end{eqnarray}
with the notation $c_{i}=\cos\alpha_{i}$ and $s_{i}=\sin\alpha_{i}$, with $i$ = 1,2,3. For simplify, we will assume $R$ is real in our calculation.

The interactions of the nine neutrino mass eigenstates, $N_{i,j}$, and charged leptons, $l_{i}$, with the gauge bosons, $W^{\pm}$ and Z, are correspondingly given by the Lagrangians:
\begin{eqnarray}
\mathcal{L}^{W^{\pm}}&=&\frac{g_2}{\sqrt{2}}U_{ij}\bar{l}_{i}\gamma^{\mu}P_L N_j W_{\mu}^-+h.c.,(i=1,...,3,j=1,...,9),\label{LFVW}
\\
\mathcal{L}^{Z}&=&\frac{g_2}{2 c_w}C_{ij}\bar{N}_{i}\gamma^{\mu}P_L N_j Z_{\mu},(i,j=1,...,9),
\label{LFVZ}
\end{eqnarray}
where $g_2$ is the coupling constant of gauge group SU(2), and $c_w$ is the cosine of the weak mixing angle. $P_{L/R} = \frac{1}{2}(1\mp\gamma_{5})$. $C_{ij}$ is defined as
\begin{eqnarray}
C_{ij}= \sum^{3}_{\alpha=1}U^{\dagger}_{i\alpha}U_{\alpha j}.
\end{eqnarray}
Here, $C_{ij}$ is also not unitary.

\section{Numerical Analysis}
\label{sec3}
To quantitatively study the non-unitary effect on various LFV processes, we perform a scan over the parameter space described as following.

Before the calculation, it is clear that present data on neutrino masses and mixing should be accounted for, which are listed in Table.\ref{Ndata}~\cite{PDG}.

\begin{table}[h]
\tbl{Neutrino oscillation data from PDG.}
{\begin{tabular}{@{}cccc@{}} \toprule
Parameter&Value&Parameter&Value\\
\colrule
$\sin^2{2\theta_{12}}$&$0.857\pm0.024$&$\Delta{m_{21}^2}$&$(7.50\pm0.20)\times10^{-5}eV^2$\\
$\sin^2{2\theta_{23}}$&$>0.95$&$|\Delta{m_{32}^2}|$&$(0.00232^{+0.00012}_{-0.00008})eV^2$\\
$\sin^2{2\theta_{13}}$&$0.098\pm0.013$&&\\ \botrule
\end{tabular}\label{Ndata} }
\end{table}
In calculation, we have randomly varied the values of $\sin^2{2\theta_{12}}$, $\sin^2{2\theta_{13}}$, $\Delta{m_{21}^2}$ and $\Delta{m_{32}^2}$ within $3\sigma$ experimental errors and set the value of $\sin^2{2\theta_{23}}$ equal to 1. The light neutrino mass spectrum is assumed to be normal ordering, i.e., $\Delta{m_{32}^2}>0$, and CP violating phases $\delta$, $\Phi_{1}$ and $\Phi_{2}$ are set to zero. The lightest neutrino mass would vary in region of $[10^{-5}, 1]$ eV. We also assume the $R$ matrix angles in Eq.(\ref{R}) are taken to be real (thus no contributions to lepton electric dipole moments are expected), and randomly vary in the range $[0, 2\pi]$. The use of $Y_{\nu}$ in Eq.(\ref{Yv}) ensures us the above neutrino oscillation data satisfied.

In SM with inverse seesaw mechanism, the relevant input parameters are the right-handed neutrino mass matrix $M_R$ and Majorana mass matrix $M_{\mu}$. Here, as mentioned before Eq.(\ref{Yv}), $M_R$ is diagonal matrix. We will make the minimal flavor violation hypothesis which consists in assuming that flavor is violated only in the standard Dirac Yukawa coupling. Under this simplification the $3\times3$ matrix $M_{\mu}$ must be also diagonal. We have randomly varied the entries of $(M_R)_{ii}$ in the range of $[1, 10^6]$ GeV and $(M_{\mu})_{ii}$ in the range of $[10^{-11}, 10^{-3}]$ GeV.

\begin{table}[h]
\tbl{Constraints used in the scan over free parameters.}
{\begin{tabular}{@{}cccc@{}} \toprule
Channel&Fraction or Limit&Channel&Fraction or Limit\\
\colrule
$W\rightarrow e\nu$&$0.1075\pm0.0013$&$W\rightarrow \mu\nu$&$0.1057\pm0.0015$\\
$W\rightarrow \tau\nu$&$0.1125\pm0.0020$&$B\rightarrow e\nu$&$<9.8\times 10^{-7}$\\
$B\rightarrow \tau\nu$&$(1.65\pm0.34)\times10^{-4}$&$B\rightarrow \mu \nu$&$<1.0\times 10^{-6}$\\
$D_s\rightarrow \mu\nu$&$(5.90\pm0.33)\times 10^{-3}$&$D_s\rightarrow e\nu$&$<1.2\times 10^{-4}$\\
$D_s\rightarrow \tau \nu$&$5.43\pm0.31\%$&$\pi\rightarrow e \nu$&$(1.230\pm0.004)\times 10^{-8}$\\
$K\rightarrow \mu\nu$&$0.6355\pm0.0011$&$\pi\rightarrow \mu \nu$&$99.98770\pm0.00004\%$\\
$K\rightarrow e \nu$&$(1.581\pm0.008)\times 10^{-5}$&$Z\rightarrow \mu \tau$&$<1.2\times 10^{-5}$\\
$Z\rightarrow e \mu$&$<1.7\times 10^{-6}$&$Z\rightarrow e \tau$&$<9.8\times 10^{-6}$\\
\botrule
\end{tabular}\label{Constraint} }
\end{table}

The experimental measurements of several rare decays should be also considered cause the parameter spaces are strongly constrained by such measurements. These rare decays have been investigated in literatures \cite{Ilakovac,Antusch,Abada,Akeroyd}. The non-unitary nature of the neutrino mixing matrix can manifest itself in tree level processes like leptonic decays of W boson and mesons ($B^+$,$D_s^+$,$K^+$ and $\pi^+$), and invisible decay of Z boson. It can also manifest in LFV decays of Z boson, LFV rare charged lepton decays like $l_i\rightarrow l_j\gamma$, $l_i\rightarrow 3l_j$, and LFV process $\mu- e$ conversion in an atom, which proceed via one loop processes, and hence can be constrained. The current experimental limits are listed in Table.\ref{BRCR} and Table.\ref{Constraint} at $1\sigma$ level. Current experimental limits are listed at the $90\%$ confidence level \cite{PDG}(except for $Z\rightarrow e^{\pm} \mu^{\pm}$, $Z\rightarrow e^{\pm} \tau^{\pm}$ and $Z\rightarrow \mu^{\pm} \tau^{\pm}$ for which the 95\% C.L bounds are given). We will use these limits to bound the parameter spaces. For the channels listed in Table.\ref{Constraint}, we require that our numerical results are compatible with the experimental values within $3\sigma$ experimental errors.

\begin{figure*}
\centering
\includegraphics[width=3.0in]{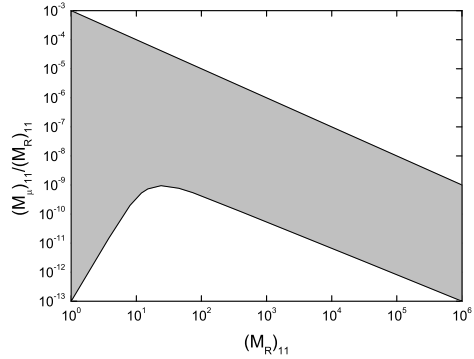}
\caption[]{Area plot of $\frac {(M_{\mu})_{11}}{(M_{R})_{11}}$ versus $(M_{R})_{11}$. The shadow region is compatible with constraints in Table.\ref{BRCR} and Table.\ref{Constraint}. }
\label{fig1}
\end{figure*}
We have studied the possible constraints on the mass matrix inputs $(M_R)_{11}$, $(M_R)_{22}$, $(M_R)_{33}$, $(M_{\mu})_{11}$, $(M_{\mu})_{22}$ and $(M_{\mu})_{33}$. It has been studied in Ref.\refcite{Deppisch} that the region of $\frac {M_{\mu}}{M_{R}}\lesssim10^{-7}$ has been excluded by considering the constraints from $\mu-e$ conversion in various nucleus, where the eigenvalues of both $M_{R}$ and $M_{\mu}$ are degenerate. We display the area plot of $\frac {(M_{\mu})_{11}}{(M_{R})_{11}}$ versus $(M_{R})_{11}$ in Fig.\ref{fig1}. It shows that, without the assumption the eigenvalues of both $M_{R}$ and $M_{\mu}$ are degenerate and by considering more constraints, the excluded region is narrowed to $\frac {M_{\mu}}{M_{R}}\lesssim10^{-9}$. In addition, the blank area in the upper right corner in Fig.\ref{fig1} is also excluded, which is not displayed in Ref.\refcite{Deppisch}. With fixed value of the scale of $M_R$ to $10^{3}$ GeV, the lower value of the scale of $M_{\mu}$ is approximately given by $M_{\mu}\sim 10^{-9}$ or $\frac {M_{\mu}}{M_{R}}\sim 10^{-12}$ in Ref.\refcite{Forero} by considering the constraint from $\mu\rightarrow e\gamma$, which is also less restrictive than our result. There are similar correlations for $\frac {(M_{\mu})_{22}}{(M_{R})_{22}}$ versus $(M_{R})_{22}$ and $\frac {(M_{\mu})_{33}}{(M_{R})_{33}}$ versus $(M_{R})_{33}$, which are not displayed to shorten the length of text.

\begin{figure*}
\centering
\includegraphics[width=3.0in]{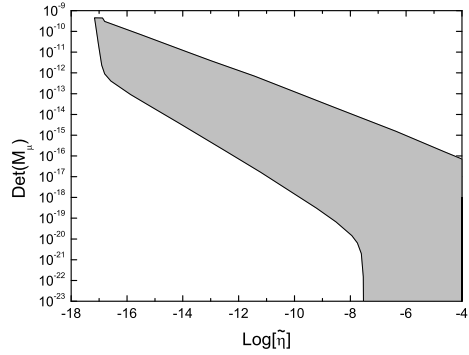}
\caption[]{Area plot of $Det (M_{\mu})$ versus $Log[\tilde{\eta}]$. The shadow region is compatible with constraints in Table.\ref{BRCR} and Table.\ref{Constraint}.}
\label{fig2}
\end{figure*}

We also investigate the dependence of $\tilde{\eta}$ on $\sin^2{2\theta_{12}}$, $\sin^2{2\theta_{13}}$, $\Delta{m_{21}^2}$, $\Delta{m_{32}^2}$, $m_{\nu_{e}}$, $(M_R)_{ii}$ and $(M_{\mu})_{ii}$. It displays $\tilde{\eta}$ strongly depends on $(M_{\mu})_{ii}$.
In Fig.\ref{fig2}, we display the determinant $Det (M_{\mu})$ versus $Log[\tilde{\eta}]$ from a scan over few $10^6$ points in parameter space in the inverse seesaw mechanism. Here, $Det (M_{\mu}) = \prod ^3_{i=1}(M_{\mu})_{ii}$. It shows that large values of unitary violation $\tilde{\eta}$ (e.g., $ 10^{-4}$) correspond to small scales of $Det (M_{\mu})$ (e.g., $10^{-15}$ \textup{Ge$V^3$}) or $(M_{\mu})_{ii}$ (e.g., $10^{-5}$ GeV). In models where lepton number is spontaneously broken by a vacuum expectation value $\langle\sigma\rangle$ \cite{Gonzalez} one has $(M_{\mu})_{ii} = (\lambda)_{ii} \langle\sigma\rangle$, where $M_{\mu}$ is diagonal as assumed. For typical Yukawas $(\lambda)_{ii} \sim 10^{-3}$ one sees that $(M_{\mu})_{ii} \sim 10^{-6}$ GeV corresponds to a scale of lepton number violation value $\langle\sigma\rangle \sim 10^{-3}$ GeV \cite{Deppisch}. Thus, if the LFV processes are observed in experiment, the vacuum expectation value $\langle\sigma\rangle$ should be the scale of $(1 - 10^{-3})$ GeV, under the assumption of typical Yukawas.

\begin{figure*}
\centering
\includegraphics[width=3.0in]{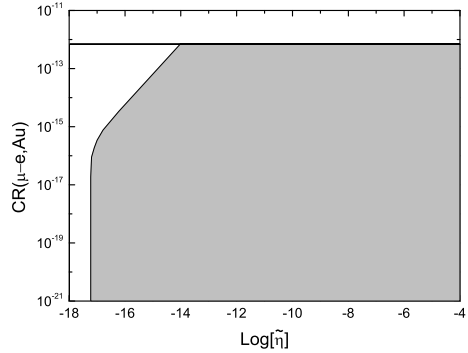}
\caption[]{Area plot of $CR(\mu-e, Au)$ versus $Log[\tilde{\eta}]$. The horizontal solid line denotes the experimental bound. The shadow region is compatible with constraints in Table.\ref{BRCR} and Table.\ref{Constraint}. }
\label{fig3}
\end{figure*}

In Fig.\ref{fig3}, we show the area plot of $CR(\mu-e, Au)$ versus $Log[\tilde{\eta}]$ in the inverse seesaw mechanism from the scan over few $10^6$ points in parameter space. The expected conversion rates $CR(\mu\rightarrow e,Au)$ are sizeable to compatible with the experimental upper limit and future experimental sensitivities in range of $10^{-14}<\tilde{\eta}<10^{-4}$. For $\tilde{\eta}<10^{-14}$, the upper limit of the $CR(\mu-e, Au)$ decreases. The expected conversion rates $CR(\mu\rightarrow e,Au)$ could be very small in the whole region of $10^{-18}<\tilde{\eta}<10^{-4}$. The area plots for $CR(\mu-e, Al)$, $CR(\mu-e, Ti)$ and $CR(\mu-e, Pb)$ versus $Log[\tilde{\eta}]$ have the same behavior.

\begin{figure}[h]
\centerline{\psfig{file=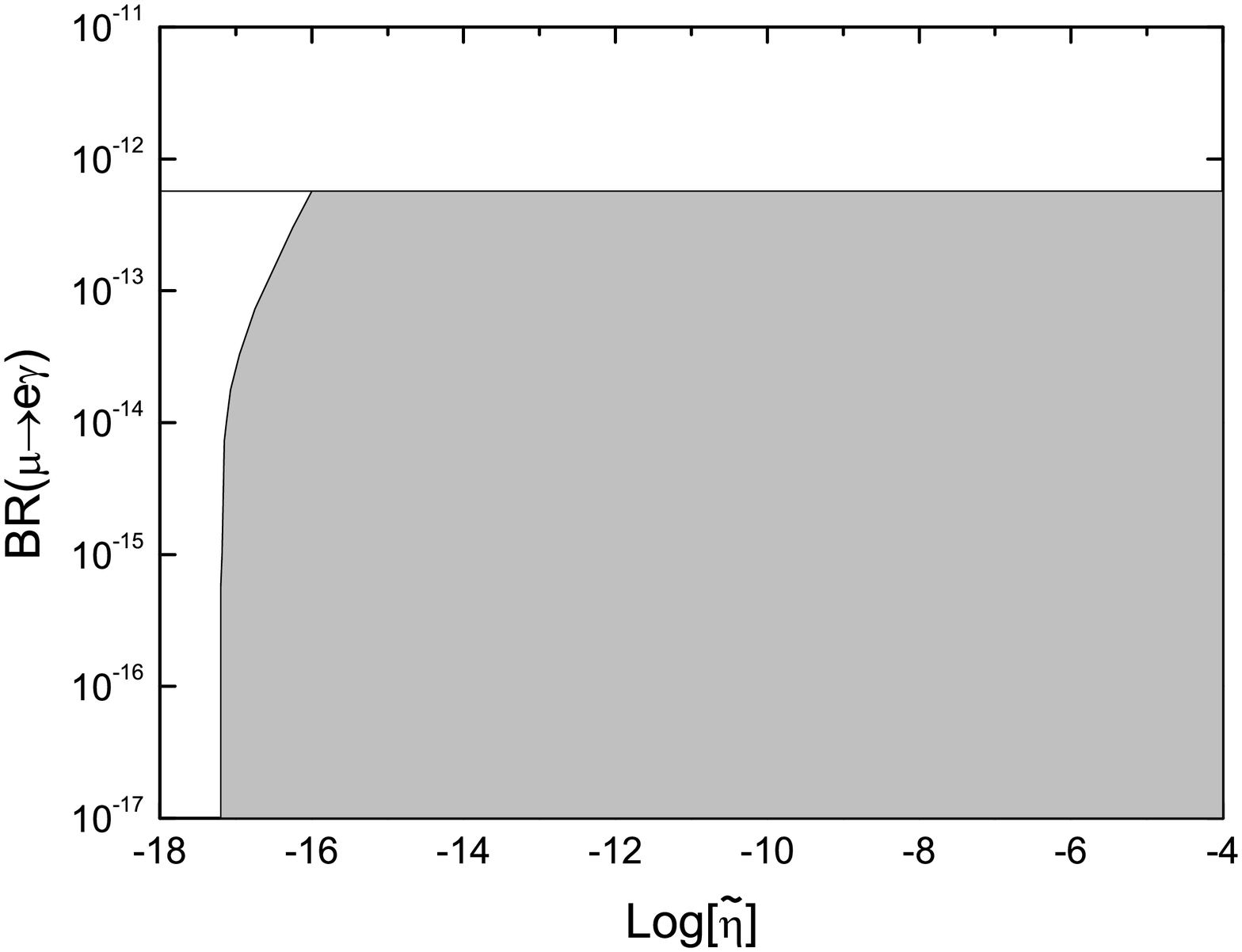,width=2.5in}\psfig{file=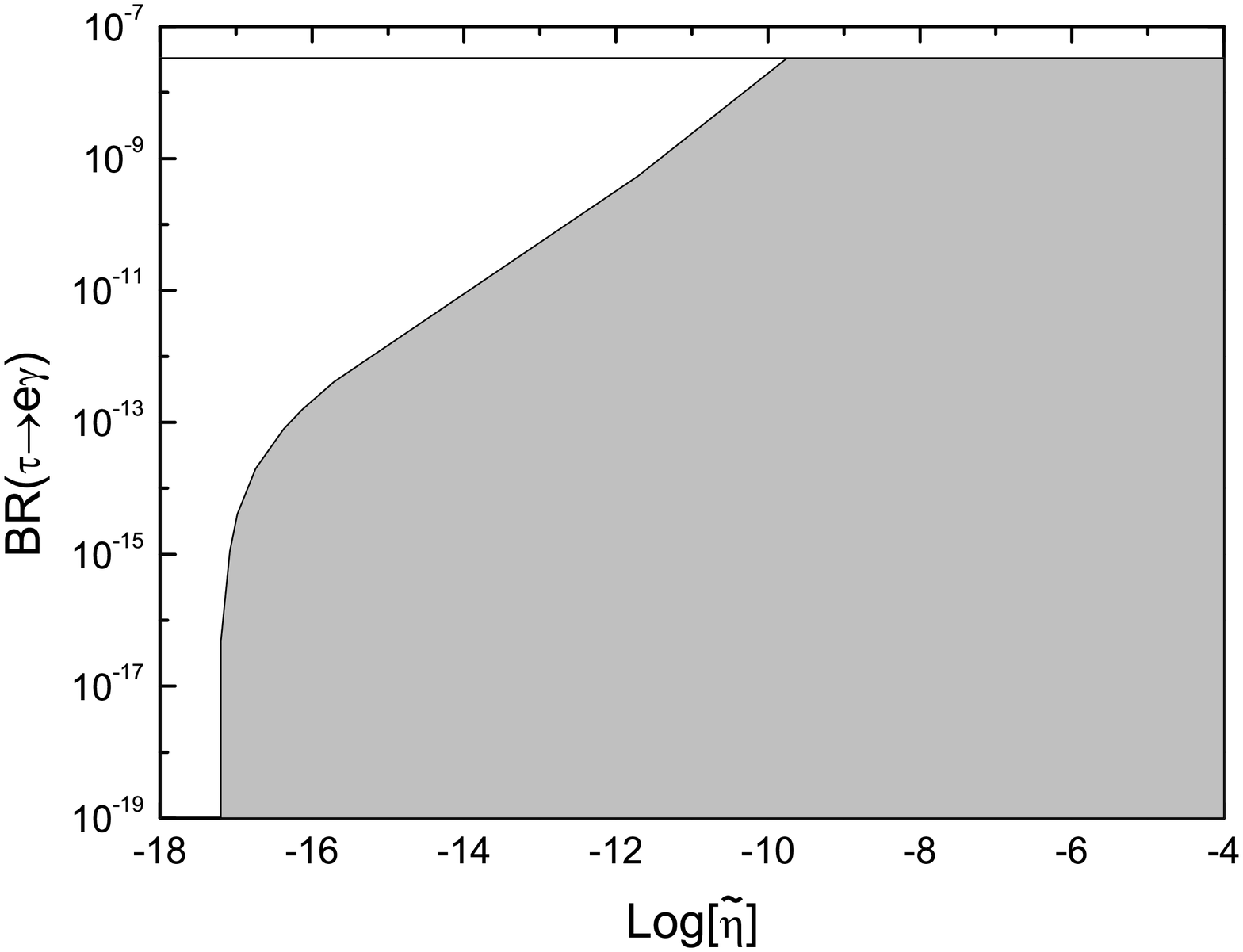,width=2.5in}}
\vspace*{8pt}
\caption{Area plot of $BR(\tau\rightarrow e\gamma)$ versus $Log[\tilde{\eta}]$ and $BR(\mu\rightarrow e\gamma)$ versus $Log[\tilde{\eta}]$. The horizontal solid line denotes the experimental bound. The shadow region is compatible with constraints in Table.\ref{BRCR} and Table.\ref{Constraint}.\protect\label{fig4}}
\end{figure}
In Fig.\ref{fig4}, we display area plot of $BR(\mu\rightarrow e\gamma)$ and $BR(\tau\rightarrow e\gamma)$ versus $Log[\tilde{\eta}]$ in the inverse seesaw mechanism from the scan over few $10^6$ points in parameter space. It shows most predictions of $BR(\mu\rightarrow e\gamma)$ are just below the experimental upper limit in range of $10^{-16}<\tilde{\eta}<10^{-4}$. In a narrow range of $10^{-18}<\tilde{\eta}<10^{-16}$, the upper limit of the predictions decreases. The prediction of $BR(\tau\rightarrow e\gamma)$ can reach to the current limits only when $\tilde{\eta}$ is large ($\tilde{\eta} > 10^{-10}$). The upper limit of the predictions decreases when $\tilde{\eta} < 10^{-10}$. It is noteworthy that $\mu\rightarrow e\gamma$ is more constraining than $\tau\rightarrow e\gamma$ in most cases from a compare between figures in Fig.\ref{fig4}. However, there is still probability that both predictions of these processes are very close to the experimental upper limit ($\tilde{\eta} > 10^{-10}$). The area plot for $BR(\tau\rightarrow \mu\gamma)$ versus $Log[\tilde{\eta}]$ has the same behavior with $BR(\tau\rightarrow e\gamma)$ versus $Log[\tilde{\eta}]$.

\begin{figure}[h]
\centerline{\psfig{file=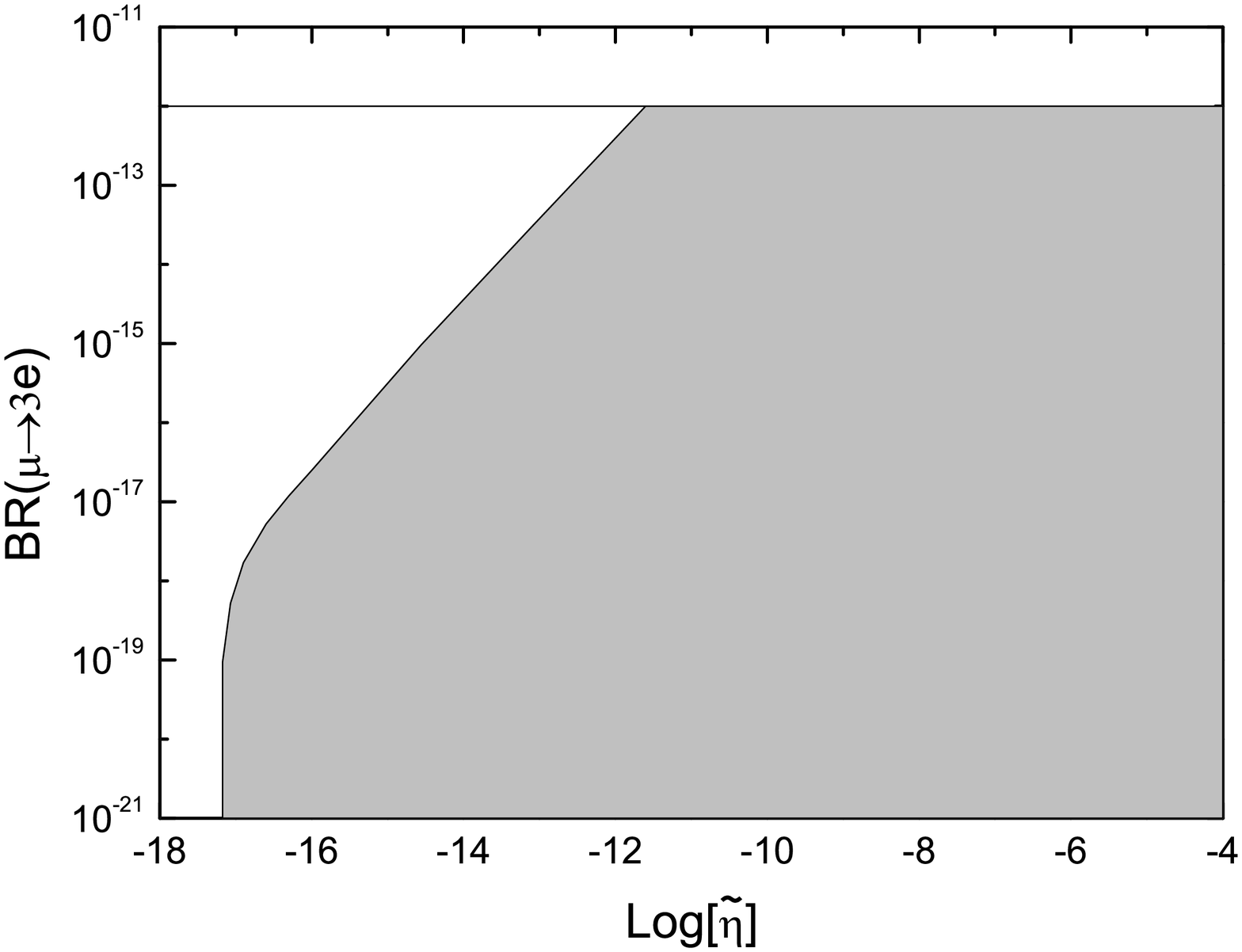,width=2.5in}\psfig{file=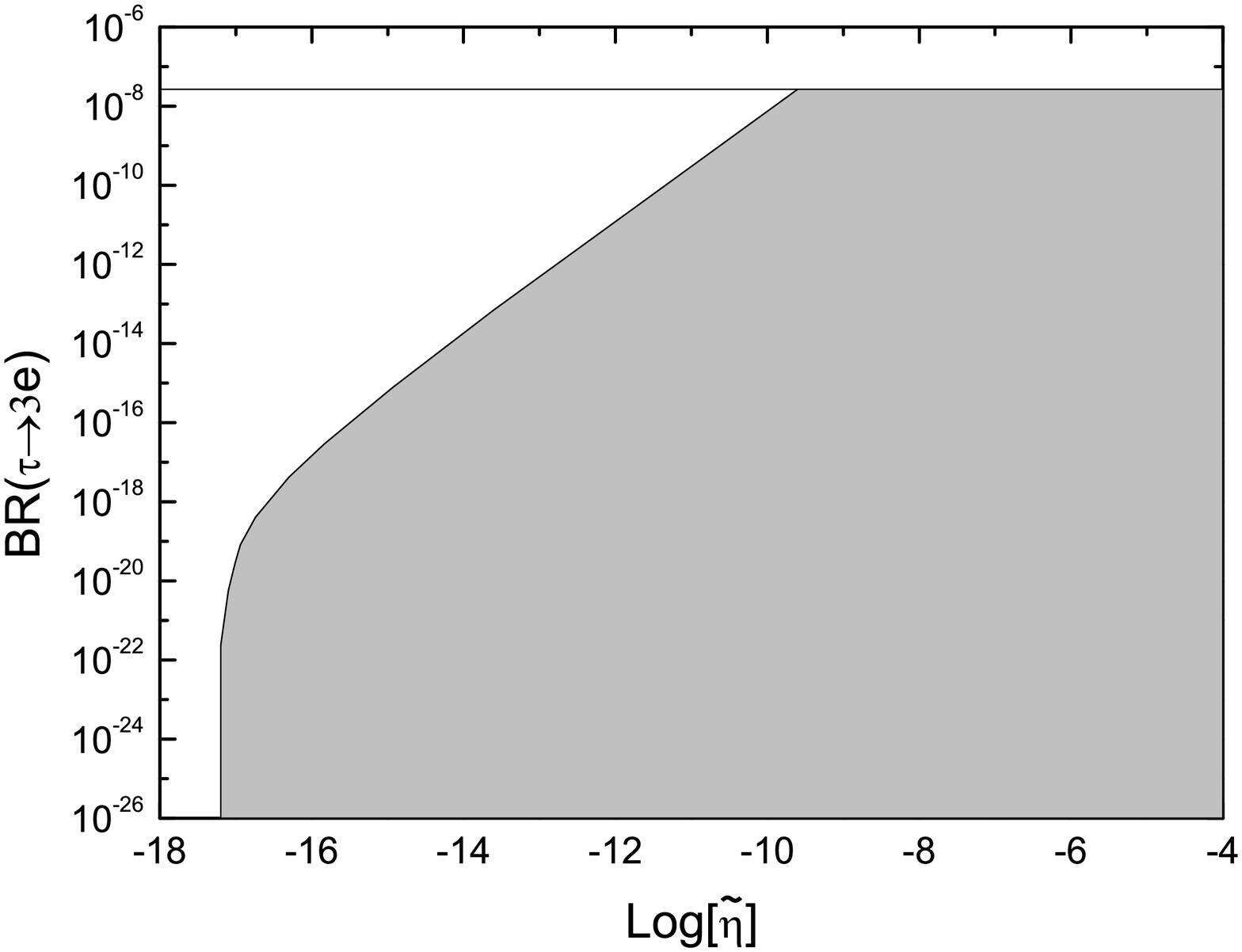,width=2.5in}}
\vspace*{8pt}
\caption{Area plot of $BR(\tau\rightarrow 3e)$ versus $Log[\tilde{\eta}]$ and $BR(\mu\rightarrow 3e)$ versus $Log[\tilde{\eta}]$. The horizontal solid line denotes the experimental bound. The shadow region is compatible with constraints in Table.\ref{BRCR} and Table.\ref{Constraint}.\protect\label{fig5}}
\end{figure}

In Fig.\ref{fig5}, we display area plot of $BR(\mu\rightarrow 3e)$ and $BR(\tau\rightarrow 3e)$ versus $Log[\tilde{\eta}]$ in the inverse seesaw mechanism. It displays that the predictions of $BR(\mu\rightarrow 3e)$ and $BR(\tau\rightarrow 3e)$ can reach the experimental limit at large values of $\tilde{\eta}$ (about $\tilde{\eta} > 10^{-12}$ and $\tilde{\eta} > 10^{-10}$). Also, the upper limits of predictions decrease when $\tilde{\eta} < 10^{-12}$ and $\tilde{\eta} < 10^{-10}$, respectively. The observation of LFV decays $\mu\rightarrow 3e$ and $\tau\rightarrow 3e$ indicate the large violation of unitary in light neutrino mixing matrix and a lower vacuum expected value $\langle\sigma\rangle $. The area plot for $BR(\tau\rightarrow 3\mu)$ versus $Log[\tilde{\eta}]$ has the same behavior with $BR(\tau\rightarrow 3e)$ versus $Log[\tilde{\eta}]$.

\section{Conclusions}
\label{sec4}
The non-unitary mixing matrix in seesaw mechanism is a generic feature for theories with mixing between neutrinos and heavy states and provides a window to probe new physics at TeV scale.

In this paper we have studied lepton flavor violation decays $l_i\rightarrow l_j\gamma$, $l_i\rightarrow 3l_j$ and $\mu-e$ conversion as a function of non-unitary parameter $\tilde{\eta}$ in the SM extended with inverse seesaw mechanism through a scan over the parameter spaces defined from the right-handed neutrino mass matrix $M_R$ and Majorana mass matrix $M_{\mu}$.
Taking account of the constraints from neutrino oscillation and various rare decays, the relevant parameter spaces are more narrow than that in Ref.\refcite{Deppisch}. The result shows that large values of unitary violation $\tilde{\eta}$ are related to small scales of $Det (M_{\mu})$ or a small vacuum expectation value $\langle\sigma\rangle$ in spontaneously lepton number broken models. In range of $10^{-14} < \tilde{\eta} < 10^{-4}$, the upper limits of predictions of $CR(\mu-e, Nucleus)$ and $BR(\mu\rightarrow e\gamma)$ can reach the sensitivity of experiment, and is promising to detect directly in experiment in near future. In range of $10^{-10} < \tilde{\eta} < 10^{-4}$, the upper limits of $BR(\tau\rightarrow e(\mu)\gamma)$, $BR(\mu\rightarrow 3e)$ and $BR(\tau\rightarrow 3e (\mu))$ can also reach the sensitivity of experiment.
Finally, searching for LFV processes can serve as a window to the new physics of seesaw nature of neutrino masses.

\section*{Acknowledgements}

The work has been supported by the National Natural Science Foundation of China (NNSFC) with Grants No.11275036 and 11047002 and Natural Science
Fund of Hebei University with Grant No. 2011JQ05, No. 2012-242.


\begin{thebibliography}{0}

\bibitem{Fukuda1}
Y. Fukuda et al., Phys. Lett. B{\bf335}(1994)237.

\bibitem{Fukuda2}
Y. Fukuda et al., Phys. Rev. Lett.{\bf81}(1998)1562.

\bibitem{Ahmad}
Q. R. Ahmad et al., Phys. Rev. Lett.{\bf89}(2002)011301.

\bibitem{Eguchi}
K. Eguchi et al., Phys. Rev. Lett.{\bf90}(2003)021802.

\bibitem{PDG}
J. Beringer et al. (Particle Data Group), Phys. Rev. D{\bf86}(2012)010001.

\bibitem{MEG2}
J. Adam et al. [MEG Collaboration], hep-ex/1303.0754.





\bibitem{Abrams}
R. J. Abrams et al. [Mu2e Collaboration], arXiv:1211.7019 [physics.ins-det].

\bibitem{Kuno}
Y. Kuno, Nucl. Phys. Proc. Suppl. {\bf225}(2012) 228.

\bibitem{Barlow}
R. J. Barlow, "The PRISM/PRIME project," Nucl. Phys. Proc. Suppl.{\bf218}(2011)44.

\bibitem{DeeMe}
M. Aoki [DeeMe Collaboration], AIP Conf. Proc. {\bf1441}(2012) 599.

\bibitem{MEG}
J. Adam et al. [MEG Collaboration], arXiv:1301.7225[physics.ins-det]..

\bibitem{SuperB}
M. Bona et al.[SuperB Collaboration].hep-ph/0709.0451.

\bibitem{BelleII}
T. Abe. et al. (Belle II).hep-ph/1011.0352.

\bibitem{Yoshida}
M. Yoshida, "The MUSIC Project," AIP Conf. Proc. {\bf1222}(2010)400.

\bibitem{Blondel}
A. Blondel et al., "Letter of Intent for an Experiment to Search for the Decay $\mu\rightarrow 3e$,"
http://www.physi.uni-heidelberg.de/Forschung/he/mu3e/, (2012).



\bibitem{Bonnet}
F. Bonnet, D. Hern$\acute{\textup {a}}$ndez, T. Ota, and W. Winter, JHEP{\bf10}(2009)076.


\bibitem{Mohapatra1}
R. N. Mohapatra and J. W. F. Valle, Phys. Rev. D{\bf34}(1986)1642.

\bibitem{Mohapatra2}
R. N. Mohapatra, Phys. Rev. Lett.{\bf56}(1986)561.

\bibitem{Ma}
E. Ma, Phys. Rev. D 80, 013013 (2009).

\bibitem{Bazzocchi}
Federica~Bazzocchi. Phys. Rev. D{\bf83}(2011)093009.

\bibitem{Bazzocchi1}
F. Bazzocchi, D. G. Cerdeno, C. Munoz, and J.W. F. Valle,
Phys. Rev. D 81, 051701 (2010).


\bibitem{Fong}
Chee Sheng Fong, Rabindra N. Mohapatra and Ilmo Sung. Phys. Lett. B {\bf 704}(2011)171.


\bibitem{Forero}
D.V. Forero, S. Morisi, M.Tortola and J.W.F. Valle.JHEP{\bf09}(2011)142.

\bibitem{Bandyopadhyay1}
ISS Physics Working Group collaboration, A. Bandyopadhyay et al.,Rep. Prog. Phys.{\bf72}(2009)106201.

\bibitem{Das}
A. Das and N. Okada, arXiv:1207.3734.

\bibitem{DeV}
P. S. B. Dev, R. Franceschini and R. N. Mohapatra.Phys.Rev. D86 (2012) 093010.

\bibitem{Bandyopadhyay2}
P. Bandyopadhyay, E. J. Chun, H. Okada and J. -C. Park.JHEP 1301 (2013) 079.

\bibitem{Fernandez}
E. Fernandez-Martinez, M. B. Gavela, J. Lopez-Pavon and O. Yasuda, Phys.Lett. B649 (2007) 427.

\bibitem{Goswami}
S. Goswami and T. Ota, Phys. Rev. D 78, 033012 (2008).


\bibitem{Antusch1}
S. Antusch, M. Blennow, E. Fernandez-Martinez and J. Lopez-Pavon,Phys. Rev. D 80, 033002 (2009).

\bibitem{Ilakovac}
A. Ilakovac and A. Pilaftsis, Nucl. Phys. B{\bf437}(1995)491.

\bibitem{Antusch}
S. Antusch, C. Biggio, E. Fernandez-Martinez, M.B. Gavela, J. Lopez-Pavon. JHEP{\bf10}(2006)084.

\bibitem{Alonso}
R. Alonso, M. Dhen, M. B. Gavela and T. Hambye,hep-ph/1209.2679.

\bibitem{Deppisch}
F. Deppisch, T. S. Kosmas, J. W. F. Valle.Nucl.Phys.B{\bf752}(2006)80.

\bibitem{Abdallah}
W. Abdallah, A. Awad, S. Khalil, H. Okada.Eur. Phys. J. C{\bf72}(2012)2108.

\bibitem{Deppisch2}
F. Deppisch and J. Valle.Phys. Rev. D{\bf72}(2005)036001.

\bibitem{Abada1}
A. Abada, D. Das, A. Vicentea and C. Weiland.JHEP{\bf09}(2011)015.

\bibitem{Abada2}
A. Abada, D. Das and C. Weiland.JHEP{\bf03}(2012)100.

\bibitem{Bhupal}
P. S. Bhupal Dev and R. N. Mohapatra.Phys. Rev. D{\bf81}(2010)013001


\bibitem{Pontecorvo1}
B.Pontecorvo, Zh. Eksp. Teor. Fiz.JETP{\bf33}(1957)549.

\bibitem{Pontecorvo2}
B.Pontecorvo, Zh. Eksp. Teor. Fiz.JETP{\bf34}(1958)247.

\bibitem{Maki}
Z.Maki, M.Nakagawa and S.Sakata, Prog. Theor. Phys.{\bf28}(1962)870.

\bibitem{Abada}
A. Abada, D. Das, A.M. Teixeira, A. Vicentea and C. Weiland.JHEP{\bf02}(2013)048.

\bibitem{Chan}
Aik Hui Chan, Hwee Boon Low, Zhi-zhong Xing. Phys. Rev. D{\bf80}(2009)073006.

\bibitem{Hooft}
G. ¡¯t Hooft, in Proceedings of 1979 Cargese Institute on Recent Developments in Gauge
Theories, edited by G. ¡¯t Hooft et al. (Plenum Press, New York, 1980), p. 135.


\bibitem{Gonzalez}
M. C. Gonzalez-Garcia, J. W. F. Valle, Phys. Lett. B{\bf216}(1989)360.

\bibitem{Casas}
J.~A.~Casas, A.~Ibarra, ~Nucl.~Phys.~B{\bf618}(2001)171.



\bibitem{Akeroyd}
A.G. Akeroyd and F. Mahmoudi.JHEP{\bf10}(2010)038.














\end{thebibliography}
\end{document}